\documentclass[preprint,showpacs,preprintnumbers,amsmath,amssymb]{revtex4}
\usepackage{graphicx}

\begin{document}

\title{Thermal conductivity of semiconductor superlattices: experimental
study of interface scattering.}
\author{J.-Y. Duquesne} 
\affiliation{Institut des NanoSciences de  Paris, UMR 7588
  \\CNRS, Universit\'e Pierre et Marie Curie - Paris 6
  \\140, rue  de Lourmel
  \\F-75005 Paris - France}

\begin{abstract}

We present thermal conductivity measurements performed in three
short-period (GaAs)$_{9}$(AlAs)$_{5}$ superlattices. The samples were
grown at different temperatures, leading to different small scale
roughness and broadening of the interfaces. The cross-plane
conductivity is measured with a differential $3\omega$ method, at room
temperature. The order of magnitude of the overall thermal
conductivity variation is consistent with existing theoretical models,
although the actual variation is smaller than expected.

\end{abstract}
 \maketitle

%%%%%%%%%%%%%%%
%Introduction
%%%%%%%%%%%%%%%
\section{Introduction}
The thermal conductivity of semiconductor superlattices is strongly
reduced with respect to the bulk values of their constituents
\cite{Yao,Lee1,Capinsky,Venkatasubramanian,Huxtable}. The heat is
mainly carried by phonons and two mechanisms can explain the effects
of the nanostructuration on thermal transport. According to the first
({\it intrinsic}) mechanism, the zone-folding involves a modification
of the phonon dispersion and allows Umklapp process for low energy
phonons \cite{Ren,Simkin}. According to the second ({\it extrinsic})
mechanism, phonons are scattered by imperfections at the interfaces
\cite{GChen1, GChen2}. Of course, both mechanisms can be active but
their relative contributions is still under debate, mainly for short
period superlattices \cite{Ward}. The assessement of the interface
role relies on discrepancies between theoretical calculations, based
on intrinsic mechanisms only, and experimental observations. In order
to take into account interfaces defects, some authors \cite{GChen2}
combined the phonon dispersion curves arising from zone folding and
interface scattering in order to fit experimental results on GaAs/AlAs
superlattices \cite{Capinsky}. Mini-Umklapp processes were not
considered and the phonon mean free path in the layers was supposed to
be identical to the corresponding bulk value. These authors derived
that 17\% of the incident phonons on an interface undergo diffuse
scattering, because of interface defects. Recently, other authors
solved the Boltzmann's equation in superlattices, beyong the constant
relaxation time approximation \cite{Ward}. They adressed the role of
the mini-Umklapp process and computed the contribution of the
intrinsic mechanism to the thermal conductivity. Comparing their
results with experimental data, they inferred that, in
(GaAs)$_{3}$(AlAs)$_{3}$, ``the reduction due to extrinsic phonon
scattering is roughly three times larger than that due to intrinsic
scattering''. Molecular dynamics calculations also stressed the
importance of interface defects \cite{Daly,Imamura}. GaAs/AlAs
superlattices were simulated with a simplified structure: the two-atom
unit cells of GaAs and AlAs are substituted for single average atoms
and the interface defects are obtained by assigning at random the
atomic sites in the last monoatomic layer of each superlattice layer
to an (average) atom or to the other, with a given probability. A 60\%
decrease of the thermal conductivity was then predicted in
(GaAs)$_{7}$(AlAs)$_{7}$ for a 50\% substitution probability
\cite{Daly,Imamura}. Lastly, the temperature dependence of the thermal
conductivity gives clues on interface scattering effects \cite{Capinsky}.

To our knowledge, no experiment has addressed directly the role of
interface imperfections. In this paper, we report on experiments
performed in (GaAs)$_{9}$(AlAs)$_{5}$ superlattices exhibiting
different interface width. We have measured their cross-plane thermal
conductivity, using the so-called differential $3\omega$ technique, at room
temperature \cite{Cahill}.

%%%%%%%%%%%%%%%%%%%%%%%%%%%%%%%%%
%Experimental set-up and results
%%%%%%%%%%%%%%%%%%%%%%%%%%%%%%%%%
% SAMPLES
\section {Experimental results}
The samples are GaAs/AlAs superlattices grown by molecular beam
epitaxy, on GaAs buffers. The nominal widths of GaAs and AlAs layers
are 2.5~nm and 1.5~nm, respectively, and the number of GaAs-AlAs
periods is 120. Table (\ref{tab:results}) displays the actual periods
determined by X-ray diffraction, versus samples.

Ideal interfaces are infinite atomically flat planes separating pure
GaAs and AlAs layers. However, actual interfaces exhibit islandlike
structures, characterized by their lateral extent and by their height
(in terms of monolayers). Direct interfaces (AlAs grown on GaAs) tend
to exhibit large islands (lateral extent from $\sim 100$ to $\sim
1000$ nm), whereas inverse interfaces (GaAs grown on AlAs) tend to
exhibit smaller islands (lateral extent $\sim 0.1$ to $\sim 10$ nm).
Compositional fluctuations inside large islands may also occur. The
quality of actual interfaces depend on a number of parameters,
including growth temperature, interruption time... and are usually
characterized either by photoluminescence \cite{Herman} or by Raman
scattering \cite{Gammon, Jusserand1}. Photoluminescence is sensitive
to large scale fluctuations, whereas Raman scattering is sensitive to
small scale fluctuations. Small scale fluctuations lead to an
effective concentration profile at the interfaces which can be
evaluated by Raman scattering and can be used to characterized the
interface width \cite{Jusserand1, Jusserand2}.

Our samples were grown under the same conditions, except for the
substrate temperature $T_{s}$, between 510 and 650~C. In that range,
small scale interface fluctuations are very sensitive to the growth
temperature \cite{Gammon}.

Indeed, the very samples that we have studied were characterized
previously by Raman scattering \cite{Jusserand1}. The intensity and
frequency shift of the Raman lines were analyzed in term of interface
broadening. Identical broadening was assumed for all interfaces. Table
(\ref{tab:results}) gives the interface width $d_{0}$, determined from
the analysis of the confined optical modes. The samples are labelled
with the same name S1, S2 and S4 as in \cite{Jusserand1}. Clearly, the
interface width increases as the growth temperature increases.

We use the differential $3\omega$ method \cite{Cahill} to measure
thermal conductivity of three superlattices, grown on GaAs substrates:
S1, S2 and S4. We compare their thermal responses with the responses
of GaAs test samples T1, T2 and T3. Previously, a thin (100~nm)
dielectric SiO$_2$ layer is sputtered on S1, S2, S4, T1, T2 and T3 in
order to insure electrical isolation between thermal transducers and
samples \cite{Lee2}. The insulating SiO$_2$ layer is deposited in a
same run on the samples and test samples. Moreover, during sputtering,
the samples and test samples are mounted on a rotating substrate
holder. In this way, the insulating layers thickness is found to be
nearly constant on T1, T2 and T3: 104, 102 and 105~nm, as measured by
ellipsometry. We extrapolate that insulating layer thickness is
constant on S1, S2, S4, T1, T2 and T3, within 1.5\%.  Thermal
transducers are processed by lift-off photolithography and thermal
evaporation of gold (200 nm) on a thin chromium adhesion layer (few
nm). The typical line width is $30~\mu m$ (S1: $35.9~\mu m$, S2:
$35.7~\mu m$, S3: $33.1~\mu m$, T1: $28.2~\mu m$, T2: $27.3~\mu m$,
T3: $28.0~\mu m$). The line length is $2.5~mm$ (S1, S2, S4) or
$3.0~mm$ (T1, T2, T3). The ratio of the heater line width to the films
thickness is around $50$. This ratio is large enough to insure
one-dimensional heat flow through the films and measurement of
cross-plane conductivities. Indeed, simulations based on \cite{Borca}
confirm that possible anisotropy of the superlattices has negligeable
effect on the cross-plane conductivity measurement.

The heater/thermometer line thermal response $T_{2\omega}$,
i.e. the in-phase and quadrature components of the temperature
oscillation at $2\omega$, reads \cite{Cahill} :
\begin{equation}
\label{eq:T2w}
T_{2\omega}=\frac{P}{l\pi\Lambda}
\int_{0}^{+\infty}
\frac{\sin^{2}y}{y^{2}\sqrt{y^{2}+iu^{2}}}dy
+\Delta T
\end{equation}
where :
\begin{eqnarray}
u&=&\sqrt{\frac{\omega}{\Omega}}\\
\Omega&=&\frac{2 \Lambda}{\rho C w^{2}}\\
\Delta T&=&\frac{P}{lw}R_{c}
\label{eq:DT}
\end{eqnarray}

At low frequency, equation (\ref{eq:T2w}) can be approximated by \cite{Lee2}:
\begin{equation}
\label{eq:T2wBF}
T_{2\omega}\simeq \frac{P}{l\pi\Lambda}
\left[
-\frac{1}{2} \ln \left(\frac{\omega}{\Omega}\right)
+K-i\frac{\pi}{4}
\right]
+\Delta T
\end{equation}

$\omega/2\pi$ is current frequency supplied to the line.
$P$ is the power supplied to the line at $2\omega$.
$\Lambda$, $\rho$ and $C$ are the thermal conductivity, mass density
and mass specific heat of the GaAs buffer.
$\Delta T$ and $R_{c}$ are the temperature drop and thermal resistance
through the insulating layer (T1, T2, T3) or through the insulating layer
and superlattice (S1, S2, S4).
$l$ and $w$ are the length and width of the heater/thermometer line.
$K \simeq 0.9066$ is a constant parameter.

%%%%%%%%%%%%%%%%%%%%%%
%Experimental results
%%%%%%%%%%%%%%%%%%%%%%

\begin{figure}
\begin{center}
\includegraphics[clip]{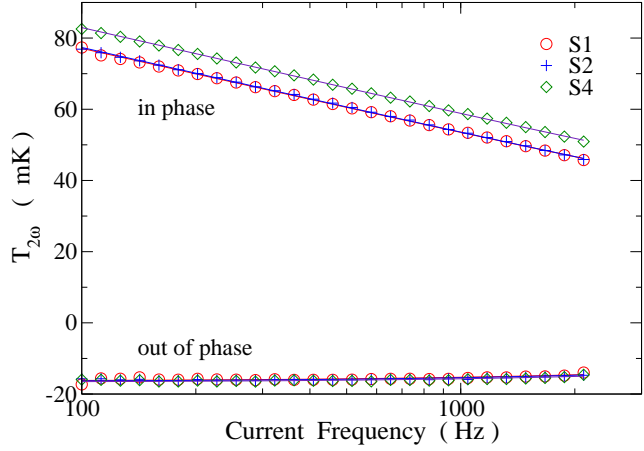}
\end{center}
\caption{\label{fig:T2w} (Color online) Typical in-phase and
  out-of-phase temperature oscillations of the heater/thermometer
  line, versus current frequency. Input power: 3~W.m$^{-1}$. Symbols:
  experimental points. Lines: fits using eq.(\ref{eq:T2w}).}
\end{figure}

\begin{figure}
\begin{center}
\includegraphics[clip]{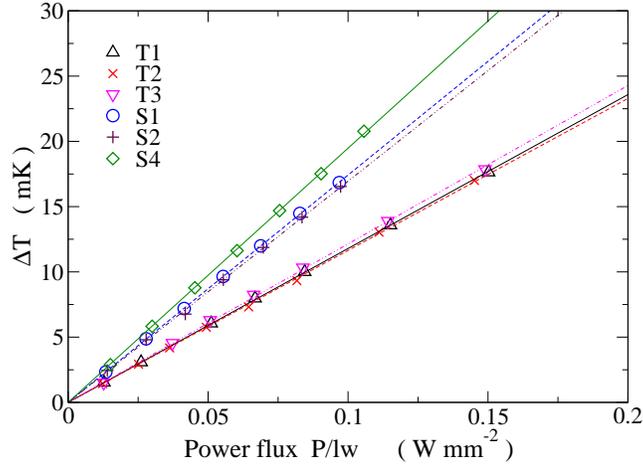}
\end{center}
\caption{\label{fig:DT} (Color online) Fitting parameter $\Delta T$
  (eq.(\ref{eq:T2w})) versus power flux through the heater/thermometer
  line. Symbols: experimental points. Lines: linear fits.}
\end{figure}

\begin{figure}
\begin{center}
\includegraphics[clip]{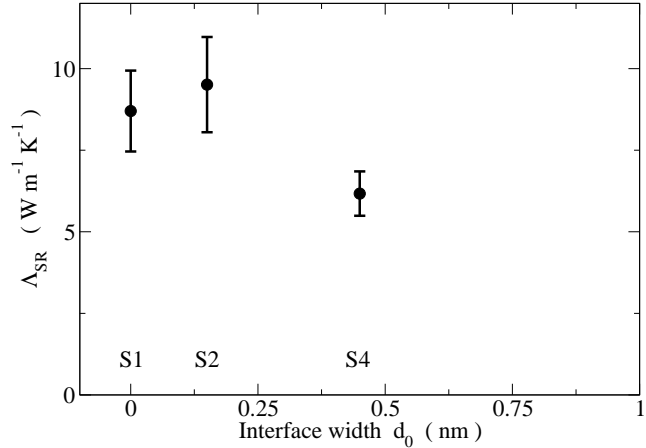}
\end{center}
\caption{\label{fig:K_SR} Thermal conductivity of
  (GaAs)$_{9}$(AlAs)$_{5}$ superlattices, versus interfaces width.}
\end{figure}

\begin{table*}
\caption{\label{tab:results} Sample parameters and experimental
  results. The superlattices were grown at a substrate temperature of
  $T_{s}$. Their period, interface width \cite{Jusserand1} and
  thickness are $d$, $d_{0}$ and $e$, respectively. $R_c$ is the
  thermal resistance through SiO$_{2}$ layers (T1, T2, T3) or through
  SiO$_{2}$ layers and superlattices (S1, S2, S4). $R_{sl}$ and
  $\Lambda_{sl}$ are the superlattice thermal resistance and thermal
  conductivity.}
\begin{center}
\begin{tabular}{cccccccc}
\hline\hline
Sample&$T_{s}$&$d$&$d_{0}$&$e$&$10^{7}\times R_{c}$&$10^{7}\times R_{sl}$&$\Lambda_{sl}$\\
&($^{\circ}C$)&($nm$)&($nm$)&($nm$)&($W^{-1}m^{2}K$)&($W^{-1}m^{2}K$)&($Wm^{-1}K^{-1})$\\
\hline
T1&&&&&1.18\\
T2&&&&&1.16\\
T3&&&&&1.21\\
S1&510&4.01&0&481.2&1.74&0.55&8.7\\
S2&550&4.01&0.15&481.2&1.69&0.50&9.5\\
S4&650&3.90&0.45&468.0&1.95&0.76&6.2\\
\hline\hline
\end{tabular}
\end{center}
\end{table*}
The thermal response of the samples and test samples is recorded at
21$^{\circ}C$, versus frequency for $P/l$ ranging from 0.5 to 3.0
$W~m^{-1}$. We fit experimental data (both in-phase and quadrature
components), using equation (\ref{eq:T2w}) and two fitting parameters:
$\Lambda$ and $\Delta T$. All the other terms are either measured
($l$, $w$, $P$) or extracted from litterature ($\rho=5317~kg~m^{-3}$,
$C=326~J~K^{-1}kg^{-1}$ \cite{Blakemore}). The uncertainty on $\rho$
and $C$ does not affect our conclusions because the comparison between
samples and test samples eliminates the substrate contribution. Figure
(\ref{fig:T2w}) displays the experimental thermal response
$T_{2\omega}$ and fits obtained at $P/l=3~W~m^{-1}$ on S1, S2 and
S4. Figure (\ref{fig:DT}) is a plot of the fitting parameter $\Delta
T$ versus the input power flux $P/(lw)$. For a given sample, the
points are aligned and, according to equation (\ref{eq:DT}), the slope
is the thermal resistance $R_{c}$. Table (\ref{tab:results})
summarizes our results.

The thermal resistance $R_{c}$ through the insulating layer and
superlattice (S1, S2, S4) is simply:
\begin{eqnarray}
R_{c}&=&R_{0}+R_{sl}\\
R_{sl}&=&\frac{e}{\Lambda_{sl}}
\end{eqnarray}
where $R_{0}$ and $R_{sl}$ are the thermal resistances of the
insulating SiO$_{2}$ layer and of the superlattice. $e$ and
$\Lambda_{sl}$ are the superlattice thickness and thermal
conductivity. In order to obtain $R_{sl}$, we use
$R_{0}=1.19~10^{-7}~W^{-1}m^{2}K$, which is the value we measure on
T1, T2, T3 within $\pm 2\%$. Table (\ref{tab:results}) summarizes our
results. Our results lie in the same range as previously reported for
GaAs/AlAs superlattices, although direct comparison is hindered by
the different periodicities of the samples \cite{Capinsky}.

The uncertainty on $R_{c}$ comes mainly from uncertainty on width $w$
of the heater/thermometer line. We estimate $\Delta w = 0.5 \mu
m$. This induces $\Delta R_{c}/R_{c}=3.2\%$. From those figures, we
deduce that layer thermal resistances are nearly equal for S1 and S2
and significantly larger on S4. Actually, we are interested in the
variation of $\Lambda_{sl}$ versus samples. As long as those
variations are concerned, $R_{0}$ variations from sample to sample
must be considered, rather than the actual values and uncertainty on
$R_{0}$. Those variations come from possible thickness variations and
are estimated to be smaller than $2\%$, in agreement with thickness
measurements on T1, T2, T3. Taking into account the uncertainty on the
thermal resistance of the layers on S1, S2, S4 and the possible
variation of the thermal resistance of the SiO$_{2}$ layers, figure
(\ref{fig:K_SR}) displays superlattice thermal conductivity variation,
versus interface width. Clearly, S1 and S2 exhibit close thermal
conductivities whereas S4 exhibits a significantly smaller thermal
conductivity.

Our data analysis neglects the boundary thermal resistances between
the various layers (line/insulating film, insulating
film/superlattice, superlattice/substrate, insulating film/substrate).
However, we may assume the boundary resistances are the same in the
various samples. In that case, their contributions are partly
cancelled out by the comparison process between samples and test
samples. Therefore, the thermal boundary effect may alter the absolute
values we derived, but not the overall behavior of the thermal
conductivity versus interface width.

%%%%%%%%%%%%%%%%%%
% Discussion
%%%%%%%%%%%%%%%%%%
\section{Discussion}

\begin{figure}
\begin{center}
\includegraphics[clip]{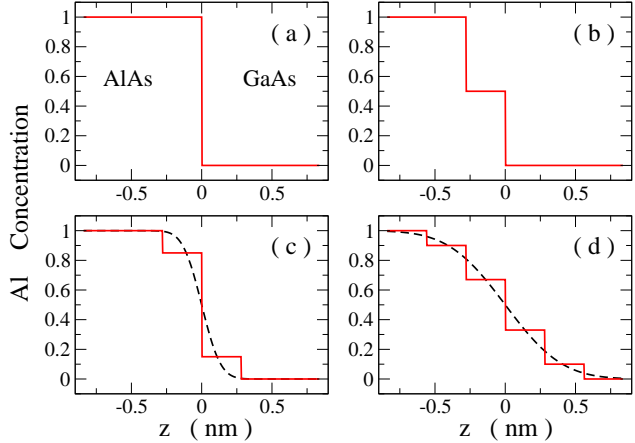}
\end{center}
\caption{\label{fig:interface} Aluminium concentration profile versus
  position along the growth axis $z$. (a) Perfect interface. (b)
  according to \cite{Daly, Imamura}. (c) $d_{0}=0.15~nm$. (d)
  $d_{0}=0.45~nm$. (c) and (d): dashed line: 'erf' profile according
  to \cite{Jusserand1}. Full line: average concentration in the
  monolayers.  }
\end{figure}

Clearly, our experiments show that S1 and S2 exhibit close thermal
conductivities whereas S4 exhibits a significantly smaller thermal
conductivity (30\% decrease).

In (GaAs)$_{9}$(AlAs)$_{5}$, the superlattice period $d$ is much
smaller than the thermal phonon mean free path $l$ in the bulk
constituents ($d=4~nm$ and $l\sim 70~nm$ \cite{GChen2}). In this so
called ``short period superlattice'' regime, we expect the interface
defects to modify significantly the phonons mean free path and, as a
consequence, the thermal conductivity. This is confirmed by molecular
dynamics calculations. Direct comparison of our experimental results
with quantitative models is difficult because existing calculations
use simplified models for both structure and interfaces \cite{ GChen2,
  Daly, Imamura}. However, we may compare the transition layer at the
interfaces involved in both models and experiments. Figure
(\ref{fig:interface}) is a sketch of the interface transition
layers. In our experiments, S1, S2 and S4 are characterized by an
interface width $d_{0}$ \cite{Jusserand1}. From $d_{0}$, we may infer
the foreign atoms concentration in the transition layer. In S2,
$d_{0}/x_{c}=0.5$ ($x_{c}=0.28~nm$ is the monolayer thickness) and the
transition layer is two monolayers in thickness: the first and last
monolayer at the interfaces. In both monolayers, the foreign atoms
concentration is 15\%. In S4, $d_{0}/x_{c}=1.6$ and the transition
layer is four monolayers in thickness: the first and last two
monolayers at the interfaces. In the monolayers the closest to the
interfaces, the foreign atoms concentration is 33\%. In the next
monolayers, the foreign atoms concentration is 10\%. In existing molecular
dynamics calculations, the transition layer is one monolayer in
thickness. It is located at the last atomic layer in each superlattice
layer and the foreign atoms concentration is 50\%. A 60\% reduction of
the thermal conductivity is then inferred in (GaAs)$_{7}$(AlAs)$_{7}$
\cite{Daly, Imamura}. On the basis of the above comparison, we
conclude that our samples lie in a range where the interface damage
should indeed have an impact on the thermal conductivity of our
samples.

Large changes of the total phonon mean free path and of the thermal
conductivity have been predicted when diffuse scattering is introduced
at the interfaces \cite{GChen2}. The largest sensitivity is achieved
for small amounts of diffuse scattering. Therefore, the conductivity
variation between S1 and S2 is surprisingly small. The conductivity
variation which is observed between S1 and S4 is consistent with the
order of magnitude predicted from molecular dynamics. However, the
30\% variation which is measured appears to be quite small with
respect the 60\% variation predicted for a less severe interface
damage.

\section{Conclusion}

We have measured the cross-plane thermal conductivity of three
short-period (GaAs)$_{9}$(AlAs)$_{5}$ superlattices, at room
temperature. Due to different temperature growth, the samples exhibit
different small scale roughness at the interfaces. The small scale
roughness is characterized by an effective interface width $d_{0}$, on
the basis of previous Raman scattering experiments. Within error bars,
no variation of thermal conductivity is observed between $d_{0}=0$ and
$d_{0}=0.15~nm$, and a 30\% decrease is observed between $d_{0}=0$ and
$d_{0}=0.45~nm$. On the basis of existing calculations, the
sensitivity of the thermal conductivity to the small scale interface
roughness seems to be not as strong as expected. Additional
experiments, as well as more realistic calculations must be performed
to clarify this point.

\section*{Acknowledgement}
I am grateful to B. Jusserand for providing samples and for fruitful
discussions.

%%%%%%%%%%%%%%%%%%%%%%%%%%%%%%%%%%%%%%%%%%%%
%References

%%%%%%%%%%%%%%%%%%%%%%%%%%%%%%
\end{document}